\documentclass[prx,aps,amssymb,twocolumn]{revtex4-1}
\usepackage{amsmath}
\usepackage{amssymb}
\usepackage{amsthm}
\usepackage{amsfonts}
\usepackage{listings}
\usepackage{enumerate}
\usepackage{latexsym}
\usepackage{color}
\usepackage[colorlinks=true,urlcolor=blue,citecolor=blue,linkcolor=blue]{hyperref}
\usepackage{psfrag}
\usepackage{ulem}

\usepackage{bm}
\usepackage{graphicx}
\usepackage{subfigure}

\newcommand{\beq}{\begin{equation}}
\newcommand{\eneq}{\end{equation}}

\newcommand{\bal}{\begin{align}}
\newcommand{\eal}{\end{align}}

\usepackage{ulem}

\input{epsf}

\begin{document}

\tolerance 10000

\newcommand{\vk}{{\bf k}}

\title{On the possibility of magnetic Weyl fermions in non-symmorphic compound PtFeSb}

\author{M.G. Vergniory$^{1,2,3}$}
\author{L. Elcoro$^{4}$}
\author{F. Orlandi$^{5}$}
\author{B. Balke$^6$}
\author{Y.-H. Chan$^{7}$}
\author{J. Nuss$^8$}
\author{A. P. Schnyder$^8$}
\author{L. M. Schoop$^{9}$}

\affiliation{${^1}$Donostia International Physics Center, P. Manuel de Lardizabal 4, 20018 Donostia-San Sebasti\'an, Spain}
\affiliation{${^2}$ IKERBASQUE, Basque Foundation for Science, Maria Diaz de Haro 3, 48013 Bilbao, Spain}
\affiliation{${^3}$ Department of Applied Physics II, Faculty of Science and Technology, University of the Basque Country UPV/EHU, Apdo. 644, 48080 Bilbao, Spain}
\affiliation{${^4}$ Condensed Matter Physics Department, Faculty of Science and Technology, University of the Basque Country UPV/EHU, Apdo. 644, 48080 Bilbao, Spain}
\affiliation{${^5}$ ISIS Neutron Pulsed Facility, Science and Technology Facilities Council,, Rutherford Appleton Laboratory, Oxford OX11 0QX, UK}
\affiliation{${^6}$ Universit\"{a}t Stuttgart, Institut f\"{u}r Materialwissenschaft - Chemische Materialsynthese, Heisenbergstr. 3
D-70569 Stuttgart, Germany}
\affiliation{${^7}$ Institute of Atomic and Molecular Sciences, Academia Sinica, Taipei 10617, Taiwan}
\affiliation{${^8}$ Max-Planck-Institut f\"{u}r Festk\"{o}rperforschung, Heisenbergstr. 1, D-70569 Stuttgart, Germany}
\affiliation{${^9}$Department of Chemistry, Princeton University, Princeton, NJ 08540, USA}

\date{\today}

\begin{abstract}
Weyl fermions are expected to exhibit exotic physical properties such as the chiral anomaly, large negative magnetoresistance or Fermi arcs. Recently a new platform to realize these fermions has been introduced based on the appearance of a three-fold band crossing at high symmetry points of certain space groups. These band crossings are composed of two linearly dispersed bands that are topologically protected by a Chern number, and a flat band with no topological charge. In this paper we present a new way of inducing two kinds of Weyl fermions, based on two- and three-fold band crossings, in the non-symmorphic magnetic material PtFeSb. By means of density functional theory calculations and group theory analysis we show that magnetic order can split a six-fold degeneracy enforced by non-symmoprhic symmetry to create three-fold or two-fold degenerate Weyl nodes. We also report on the synthesis of a related phase potentially containing two-fold degenerate magnetic Weyl points and extend our group theory analysis to that phase. 
This is the first study showing that magnetic ordering has the potential to generate new threefold degenerate Weyl nodes,
 advancing the understanding of magnetic interactions in topological materials. 
\end{abstract}

\maketitle


\section{Introduction}

In the presence of spatial inversion or time-reversal symmetry (TRS), Weyl fermions can appear in condensed matter as topologically protected electronic band crossings  close to the Fermi level with quantized Chern numbers. First theoretically predicted~\cite{PhysRevB.83.205101,PhysRevX.5.011029,NatCommHuang} and then experimentally discovered~\cite{PhysRevX.5.031013,Xu07082015} in the family of TaAs compounds, they display exotic transport properties and spectroscopic phenomena such as the chiral anomaly,  large negative magnetoresistance with values exceeding the ones achieved in
semiconductors and metals, and surface disconnected Fermi arcs\cite{Nielsen1983389,PhysRevB.83.205101,Xiong413,PhysRevX.5.031023,PhysRevB.90.155316,PhysRevLett.114.206401,PhysRevB.91.165105,PhysRevB.94.241405}.


As promising new platforms for novel applications in spintronics, several families of materials have been proposed as candidate materials for Weyl semimetals (WSMs) \cite{soluyanov2015nature,hasanatcomm,PhysRevB.94.161116,PhysRevB.94.161401,TQC,Schooprev}. However, when time-reversal symmetry is preserved,  Weyl nodes appear in multiples of $4$, resulting in complicated band structures challenging their integration into spintronic devices.\\

If TRS is broken, the number of Weyl nodes is reduced by a factor of two, which is why the search for magnetic WSMs is crucial for realizing possible applications.  Unfortunately, materials proposed theoretically as magnetic Weyl fermions\cite{xu2011,burkov2011,liuweyl2014,OngNatMat,PhysRevLett.117.236401,FelserWeylsemimetal} have not yet unambiguously been experimentally confirmed, and a good candidate material is still needed. Further, in order to search for material candidates efficiently, it is important to fully understand when Weyl nodes can appear in a material.
Recently, it has been proposed that topological band crossings protected by non-symmmorphic symmetries will result in fermions beyond Weyls\cite{PhysRevB.85.155118,PhysRevLett.116.186402,Bradlynaaf5037}; these fermions have also been called {\it New Fermions}. They are based on the appearance of three-, six- or eight-fold band degeneracies  that are located at high-symmetry points in several space groups (SGs) and are enforced by group theory, thus appearing in all materials that crystallize in these SGs. The challenge is to find a material where these band crossings appear close to the Fermi level\cite{ToppNJP}. In particular, it was shown that the groups $I2_{1}31'$ (No 199.13) and $I4_{1}321'$ (No. 214.68), host a three dimensional band crossing at the $P$ point, which is a time-reversal (TR), non-invariant point of the Brillouin zone (BZ). These three-fold degenerate new fermions can be viewed as a Weyl nodes with a nominal topological charge of 2 intersected by an additional flat band with no charge. However, the presence of these three-fold fermions in real materials is very rare. The {\it International Crystallographic Database} (ICSD)\cite{ICSD} does not contain many compounds that crystallize in the two parent grey groups of these two groups, and of the known ones, many compounds host the three-fold band crossings far away from the Fermi level. This is the reason why we introduce a different approach for realizing three-fold degenerate fermions in this paper.
In the presence of TR symmetry, six space groups can host six-fold degeneracies, which are in general much more abundantly found in real materials, since the amount of materials crytallizing in these space groups is much larger\cite{ICSD}. One space group hosting six-fold  band degenercies is  $P2_{1}31'$ (198.10)\cite{Bradlynaaf5037}, which is a folded version of space group $I2_{1}31'$. It corresponds to a simple-cubic Bravais lattice, and the six-fold degeneracy occurs at the TR invariant R point , which is located at the corner of the BZ. Moreover, the magnetic $P2_{1}31'$, is one of the Sohncke groups, making materials crystallizing in this space group promising
for experimental applications such as the observation of the quantized photogalvanic 
effect\cite{PhysRevLett.116.077201,FerNatCom,PhysRevLett.119.206401,PhysRevB.94.241105}.

It was recently proven that band degeneracies that are a result of non-symmorphic symmetries are affected by magnetic order\cite{Schoopeaar2317,TOPP2017,Canoprep}.  In particular, TR symmetry breaking can reduce the band degeneracy. Thus, breaking TR symmetry in a compound containing a six-fold degenerate point, could generate magnetic WSMs based on  two or three-fold band crossings. So far, three-fold magnetic Weyl fermions have not been described either theoretically or experimentally, to the best of our knowledge.

\begin{figure}
\centering
\includegraphics[scale=0.3]{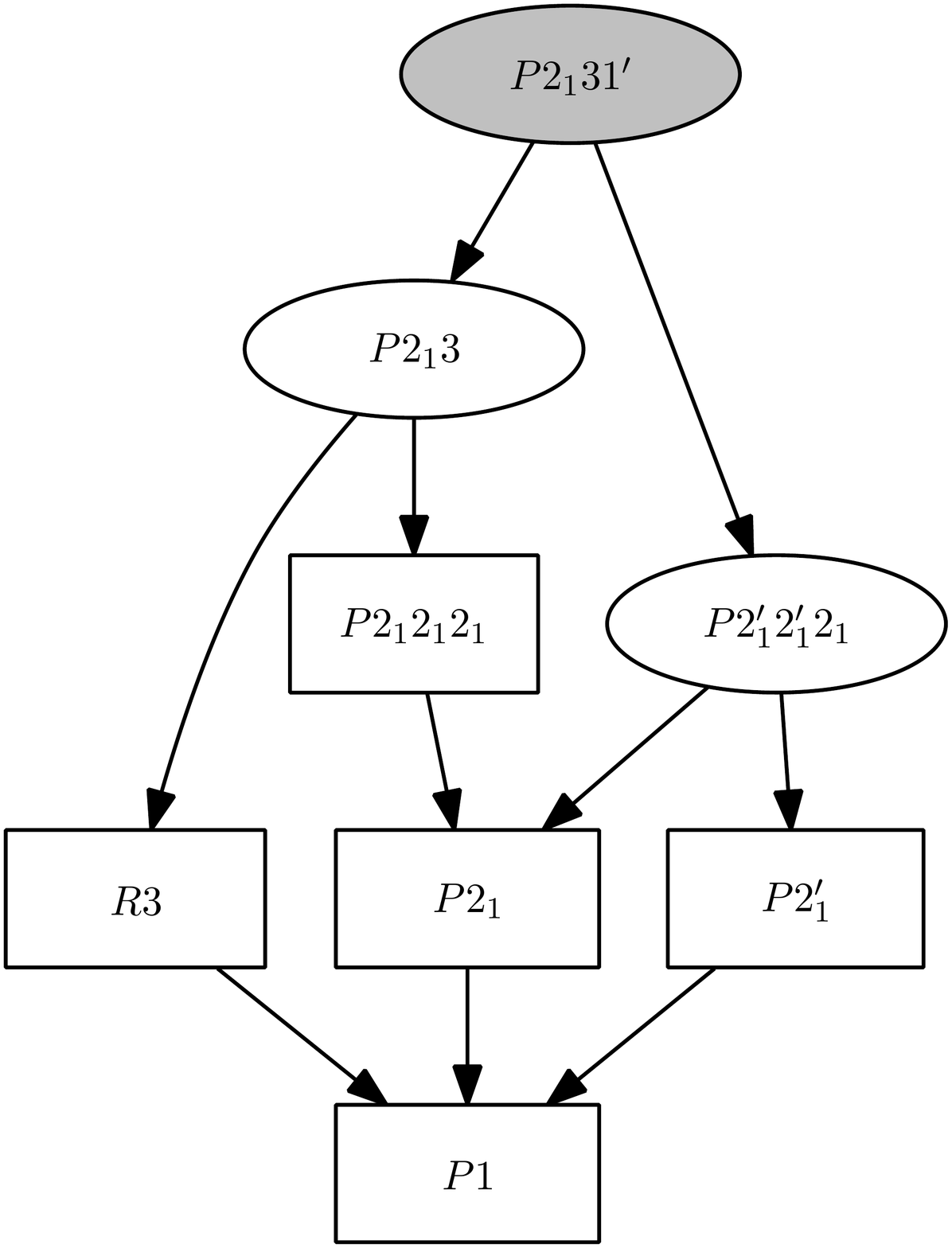}
\caption{Graph of magnetic subgroups of the gray group $P2_{1}31'$ (No.198.10) compatible with no change of the unit cell and a non-zero component of the magnetic moment at the 4a wyckoff position of the Fe atom in the PtFeSb compound. The subgroups and the graph have been obtained from the tool {\it k-Subgroupsmag}\cite{KSUB1,KSUB2}.} 
\label{fig0}
\end{figure}

In this paper, we investigate the compound PtFeSb, which was reported to crystallize in the non-symmorphic SG $P2_{1}31'$\cite{FePtSb} as candidate to host magnetic Weyl nodes. By means of  first-principle calculations we will introduce the different topological magnetic phases this compound can display. Symmetry analysis based on magnetic group theory will solidify our $ab-initio$ claims of the existence of magnetic Weyl nodes at high symmetry points. Finally, we will present experimental progress on the characterization of this compound.
Unlike previously reported \cite{FePtSb}, we were only able to synthesize a material that crystallizes in space group $Pa\bar 31'$ (No. 205.34), adopting a disordered structure compared to the reported one. We therefore extend our theoretical analysis to $Pa\bar 31'$, finding that similar conditions apply for the enforced band degeneracies, finding that they can split to form two-fold Weyl nodes.

\begin{figure*}
\centering
\includegraphics[scale=0.6]{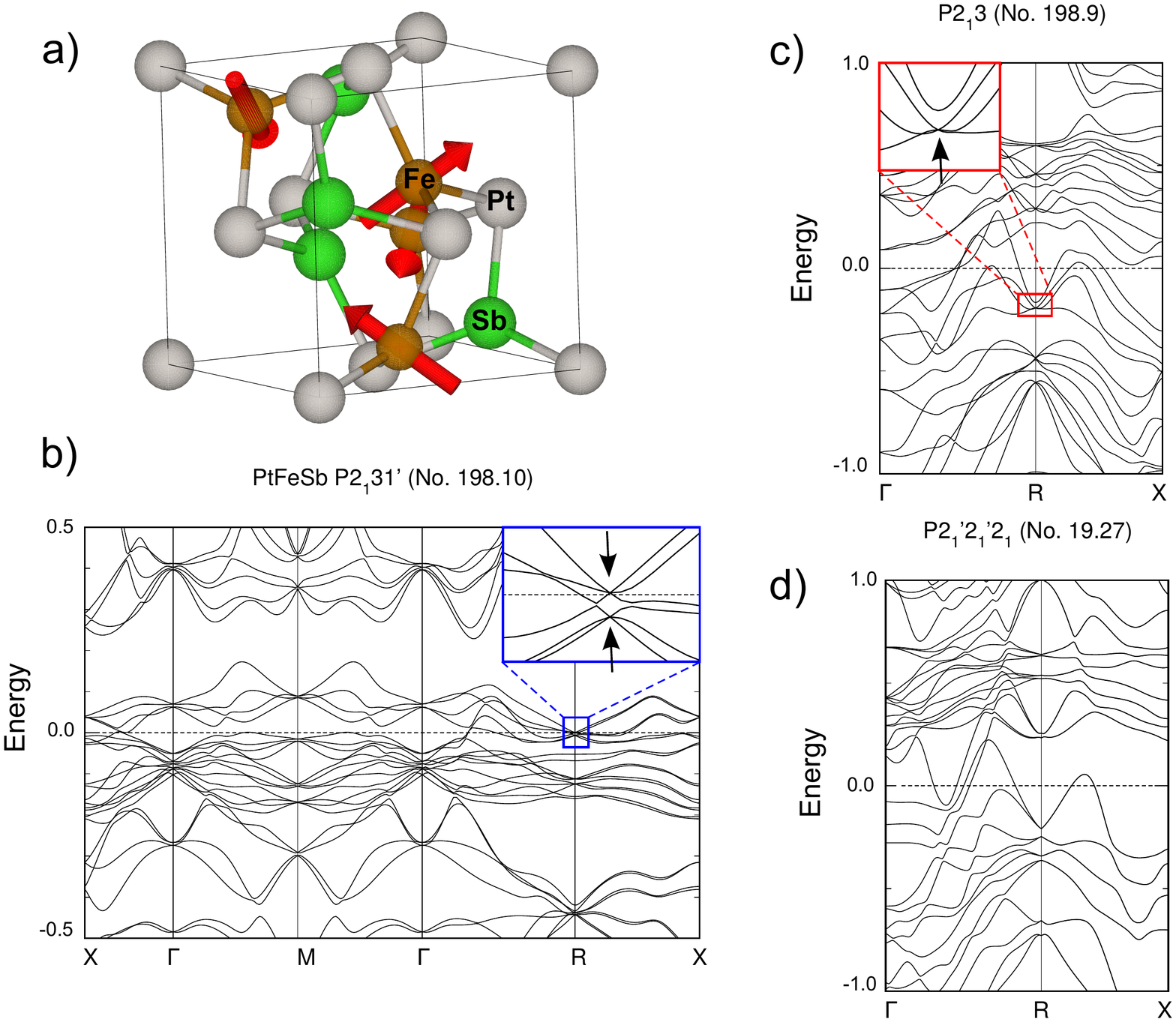}
\caption{(a) Crystal structure of PtFeSb with spins aligned so that it adopts the $P2_{1}3$ (No. 198.9 in the BNS setting) magnetic group. Fe, Pt and Sb atoms are shown in brown, grey and green respectively. (b)  Paramagnetic bulk band structure of PtFeSb along high symmetry lines including spin-orbit coupling, six-fold band crossings are highlighted. (c) Bulk band structure of PtFeSb in the magnetic chiral structure $P2_{1}3$, the inset shows three-fold and one-fold degenerate bands. (d) Bulk band structure of PtFeSb in a ferromagnetic phase with the magnetic moments aligned along (001) direction (magnetic group $P2_{1}'2_{1}'2_{1}$ No. 19.27).} 
\label{fig1}
\end{figure*}

\section{Method of Calculation}

The first-principles calculations have been performed within the DFT\cite{Hohenberg-PR64,Kohn-PR65} as implemented in the Vienna Ab initio Simulation Package (VASP)\cite{Kresse199615,PhysRevB.48.13115}. The interaction between
ion cores and valence electrons was treated by the projector augmented-wave method\cite{vaspPaw} and the generalized gradient approximation (GGA) for the exchange-correlation potential with the
Perdew-Burke-Ernkzerhof for solids (PBEsol) parametrization~\cite{PBE}. In order to account for the magnetic character of
the system we performed spin-polarized calculations. Spin-orbit coupling (SOC) was included in the Hamiltonian by adding scalar relativistic corrections using the second variation method\cite{PhysRevB.62.11556}. A Monkhorst-Pack
k-point grid of (4$\times$4$\times$4) for reciprocal space integration and 600 eV energy cutoff of the plane-wave expansion have been used to get a residual error on the forces of less than 1 meV/\AA, resulting in a fully converged electronic structure including SOC.

The matrices of the irreducible representations (irreps) of the (double) space groups have been obtained from the tool {\it representations DSG}\cite{Elcorodsg,PhysRevE.96.023310} in the {\it Bilbao Crystallographic Server}\cite{BCS1,BCS2,BCS3}, and the dimensions of the irreducible co-representations of the (double) magnetic groups have been inferred from the dimensions of these irreps. For completeness we have also used  {\it ISODISTORT} software\cite{ISODISTORT}.


\section{Experimental Details}
Samples were prepared by placing elemental Pt, Fe and Sb in a molar ratio of 1:1:1.5 in sealed quartz tubes. The tubes were then heated to 1150 $^\circ$ C  for 24 hours. They were subsequently cooled with a rate on 1.5$^\circ$/h to 700$^\circ$ C and then quenched into ice water. Without quenching, multi-phase samples were obtained. Without a large excess of Sb, an impurity phase, which could be identified as a Pt-Fe alloy appeared in the x-ray diffraction pattern. Powder x-ray diffraction measurements were performed at room temperature on a Bruker D8-Advance with Cu-K$\alpha_1$ radiation (Ge(111) monochromator, $\lambda$ = 1.54059 \AA), in reflection geometry, using a Vantec detector. High temperature magnetic measurements were performed on an MPMS from Quantum Design, equipped with a high temperature furnace.

\section{Results}
\subsection{Theoretical prediction} 


PtFeSb was originally reported to crystallize in the space group $P2_{1}31'$ (No. 198.10)\cite{FePtSb}. It was expected to crystallize in the half-Heusler crystal structure similarly to related compounds, but was found to exhibit a closely related structure in which Sb and Fe are displaced from the ideal (half-Heusler) positions, breaking all the mirrors and inversion symmtries (see Fig.~\ref{fig1}(a))\cite{xtal-str,PhysRevB.56.13012}. The structure can be viewed as an ordered version of the pyrite crystal structure. Pyrites crystallize in SG $Pa\bar 31'$ (No. 205.34) and have the nominal composition of \textit{MX$_2$}, where the \textit{M} atoms sit on the Wyckoff position 4a and the \textit{X} atoms on 8c. In the ordered version of the structure, the 8c position splits into two 4a positions in SG  $P2_{1}31'$ (No. 198.10), thus forming an ordered ternary structure type.
SG  $P2_{1}31'$ has four time-reversal invariant momenta (TRIMs), $\Gamma$, $R$, $M$ and $X$. We first elucidate the band topology in the absence of  magnetism. At the corner of the BZ, at R=($\pi$,$\pi$,$\pi$), 3 generators leave this point invariant, $C_{3,111}^{-1}$ along the (111) axis, $C_{2x}$ and $C_{2y}$; adding TRS the little group can be described by a six- or a two-dimensional irreducible representation (irrep)\cite{Bradlynaaf5037}. The six dimensional irrep is visible as the six-fold degenerate band crossing in Fig.~\ref{fig1}(b)), above and below Fermi level. These chiral six-fold bands near the Fermi level should feature a spin-1 Weyl fermion. In addition, a four-fold-degenerate point appears at $\Gamma$, with a Chern number of $\pm$4\cite{PhysRevLett.115.036806,PhysRevLett.119.206402}.  The two TRIMs are unrelated by symmetry and they are therefore able to display gigantic Fermi arcs. These Fermi arcs should be easily detectable with spectroscopic methods. Since we have a non-zero topological charge when TRS is preserved, we expect the six-fold crossing to split in Weyl nodes when TRS breaks. 
\begin{table}[h!]
\begin{center}
\caption{Possible dimensions of the irreps at R point of the ferromagnetic and magnetic chiral phases of PtFeSb}
\label{posi}
\begin{tabular}{ccc}
\hline
\hline
Magnetic Group & K point & Dim \\
 \hline
 $P2_{1}3$ (No. 198.9)  & R & 1,3\\
 $P2_{1}'2_{1}'2_{1}$ (No. 19.27) & R & 2 \\
 
\hline
\hline
\end{tabular}
\end{center}
\end{table}
Breaking TRS can give rise to different magnetic groups (of type I and III) without changing the unit cell. Fig.~\ref{fig0}  shows the subgroup-tree compatible with a non-zero magnetic moment at the position occupied by the Fe atoms (4a). Only the magnetic groups $P2_{1}3$ and $P2_{1}'2_{1}'2_{1}$ are maximal subgroups. In this work, we are going to focus on the $P2_{1}3$ (No. 198.9) and $P2_{1}'2_{1}'2_{1}$ (No. 19.27) subgroups, as they can hold three and two dimensional co-representations at point $R$, respectively (see Table~\ref{posi}). In all other magnetic groups, magnetic order will cause the bands to split fully, so that they are no longer degenerate.

{\it Possible three-fold magnetic Weyl fermions} The group $P2_{1}3$ (No. 198.9) holds a magnetic phase that breaks TRS while keeping the cubic structure. In this way the six-dimensional irrep ${\bar R_{7}}{\bar R_{7}}$, splits into two ${\bar R_{7}}$ irreps of three dimensions.  Accordingly, the two dimensional ${\bar R_{4}}{\bar R_{4}}$ and ${\bar R_{5}}{\bar R_{6}}$ irreps split into 1 dimensional ones (${\bar R_{4}}$,${\bar R_{5}}$ and ${\bar R_{6}}$).  In order to maintain the cubic symmetry, there is only one possible magnectic configuration: the direction of the magnetic moments needs to be along (111) cubic diagonals and the sum of the magnetic moment of the Fe atoms needs to be 0 ($\sum{\bf m}_{Fe}=0$). This complex non-collinear magnetic structure (see Fig.~\ref{fig1}(a)) has been reported in other compounds crystallizing in the same SG\cite{PhysRevB.69.054422}. As predicted by magnetic group theory, the band splittings are visible in our ab-initio calculations, shown in Fig.~\ref{fig1}(c). We preformed this calculation for different values of Hubbard-U (0, 1, 2 and 3 eV) obtaining the same value of the magnetic moment m=1.7 $\mu_{B}$.



{\it Possible two-fold magnetic Weyl fermions} $P2_{1}'2_{1}'2_{1}$ is another maximal subgroup of the magnetic group $P2_{1}31'$ that can hold a magnetic phase with three different components of the magnetic moment (m$_x$,m$_y$,m$_z$) on the Fe atoms, symmetry imposes that the sum of the 4 magnetic components on the $x$ and $y$ axis needs to be 0 independently ($\sum{\rm m}_{x}=0$ and $\sum{\rm m}_{y}=0$). In this magnetic phase the co-representation ${\bar R_{7}}{\bar R_{7}}$ of the grey SG $P2_{1}3$ (see Fig. \ref{fig0}),  splits into three 2-dimensional ones\cite{Miller}. We performed our band structure analysis considering a magnetization along the (001) direction. The band splitting predicted by group theory can again be observed in the calculated band structure; in Fig.~\ref{fig1}(d) the magnetic two-fold Weyls nodes are visible at the $R$ point.

\subsection{Experimental data} 
\begin{figure}[h!]
\centering
\includegraphics[scale=0.35]{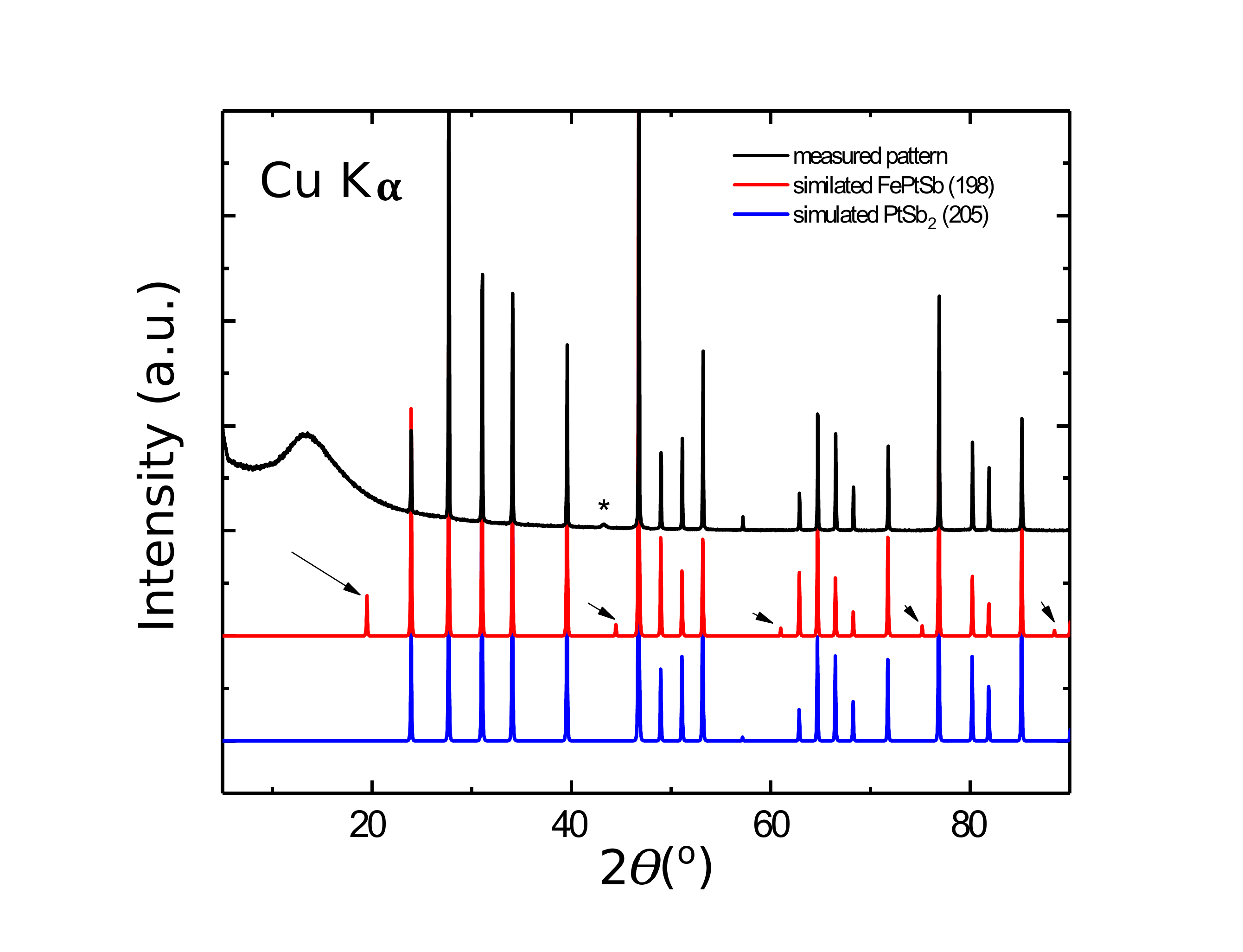}
\caption{Powder x-ray diffraction of  PtFe$_{1.1}$Sb$_{1.33}$. Measured data are shown in black and simulated patterns for PtFeSb and PtSb$_2$ are shown in red and blue respectively. Peaks that can serve as an indicator for the ordered phase are indicated by arrows. None of these are observed in the experimental data. The impurity phase is marked with an asterisks .}
\label{XRD}
\end{figure}

Motivated by our theoretical analysis, we tried to synthesize the material to explore the magnetic structure, In order to obtain information about the magnetic groups  accessible in PtFeSb.
Phase pure samples could only been obtained if about 50 \% excess of Sb was used during the synthesis. Fig. \ref{XRD} shows the powder x-ray diffraction pattern of a sample obtained with the starting composition of PtFeSb$_{1.5}$. A very small amount of impurity phase is present, which is marked with an asterisks. If different starting compositions are used, this peak increases significantly in intensity. It's position roughly matches the database entry of a bcc-type Pt-Fe alloy \cite{ICSD}. In comparison to the remaining reflections, the impurity reflection always appeared much broader, indicating  that the Pt-Fe alloy side-phase can adopt a range of composition. Chemical analysis of the sample with the smallest amount of impurity revealed a composition of PtFe$_{1.1}$Sb$_{1.33}$. Fig. \ref{XRD} also displays the simulated x-ray patterns for PtFeSb \cite{FePtSb} and PtSb$_2$ \cite{PtSb2}. Both materials exhibit the exact same lattice constant, hence doping between the two end members of PtFe$_{x}$Sb$_{2-x}$ would not cause a peak shift. However, a clear distinction between the two space groups of the ordered ternary or of a disordered pyrite structure can be made, since the former one requires superstructure  reflections. These extra reflections are indicated with arrows in Fig. \ref{XRD}. It is very obvious that the sample does not exhibit these reflections, clearly indicating that it is adopting a disordered version of the pyrite crystal structure in space group $Pa\bar 31'$. If we would assume that Fe would exclusively substitute Sb, the measured composition would not match the expected MX$_2$ composition of a pyrite compound. We thus assume that Fe can substitute on both, the Pt and the Sb site. We could assume a nominal composition of (Pt$_{0.9}$Fe$_{0.1})($Sb$_{1.3}$Fe$_{0.7}$) for the material, which would match measured composition in a very disordered pyrite phase, in agreement with our powder x-ray diffraction analysis. We tried to confirm this model with single crystal diffraction. Although we were able to obtain single crystal diffraction data of reasonable quality, we were unable to reliably refine the data in respect to the location of the Fe atoms. Single crystal diffraction did however confirm the space group of the material to be $Pa\bar 31'$. Since no major impurity phases were present and since the reflections, both in powder and single crystal diffraction are sharp, we conclude the we made a highly disordered pyrite phase with the chemical composition PtFe$_{1.1}$Sb$_{1.33}$. Since there is no change in lattice constant between PtSb$_2$ and PtFeSb, we cannot be sure that the material synthesized here has a range of intermediate compositions. We can however certainly say that the pyrite phase PtSb$_2$ can be doped with a significant amount of Fe.

\begin{figure}[h!]
\centering
\includegraphics[scale=0.35]{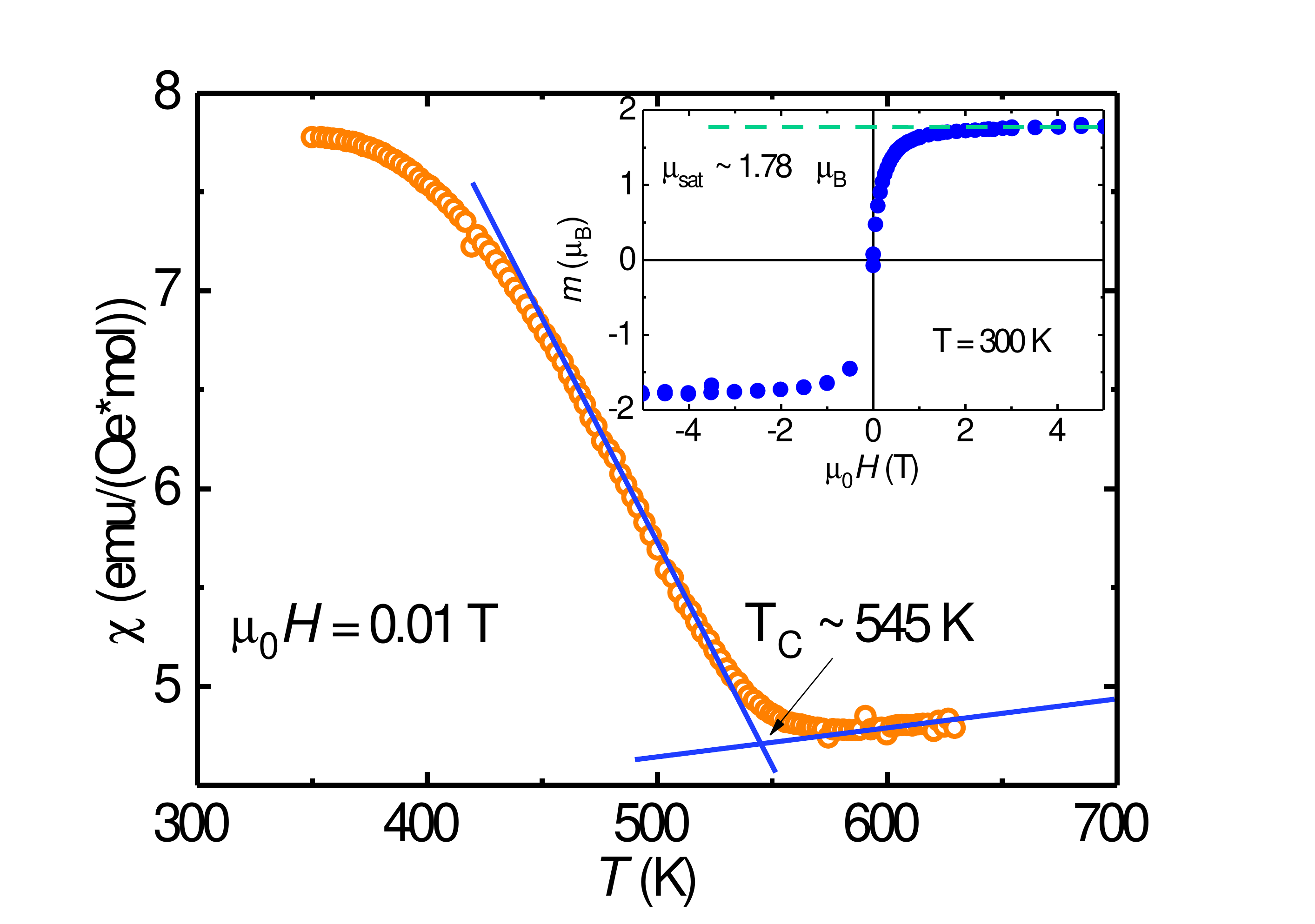}
\caption{Magnetic properties of  PtFe$_{1.1}$Sb$_{1.33}$. The Temperature dependent susceptibility is shown in the main panel. While a transition from ferromagnetic to paramagnetic order is observed as roughly 545 K, the susceptibility remains high, indication that a fraction of the sample remains ferromagnetic above T$_C$. The inset shows the field dependent magnetic moment measured at T=300K. The saturated moment is too high for being cause by an impurity phase.}
\label{mag}
\end{figure}

We further analyzed the magnetic properties of our sample. We found that the sample sticks to a simple bar magnet at room temperature, indicating ferromagnetic behavior. Fig. \ref{mag} shows the temperature dependent susceptibility of PtFe$_{1.1}$Sb$_{1.33}$ between 300 K and 630 K. While we observe a drop in the susceptibility at around 545 K, which indicates a transition form ferromagnetic to paramagnetic order, the susceptibility remains relatively large above the transition. This suggests that a portion of the sample remains ferromagnetic above the transition, supporting our previous discussion of a possible range of compositions in the sample. Nonetheless, we can exclude elemental Fe as the cause for the ferromagnetic behavior due to two reasons. For once, the saturated magnetic moment at 300 K is too high for just being caused by an Fe impurity, which is not visible in the powder x-ray diffraction pattern and could thus not be larger than 5 \% of the sample. Secondly, the Curie temperature of Fe is much higher, about 1043 K \cite{Spaldin}.We can therefore conclude that we could synthesize a ferromagnetic pyrite phase in group $Pa\bar 31'$. This is not surprising, since $P2_{1}31'$ is a subspace-group of $Pa\bar 31'$ that lacks inversion symmetry.

\begin{figure*}[h!]
\centering
\includegraphics[scale=0.6,angle=90]{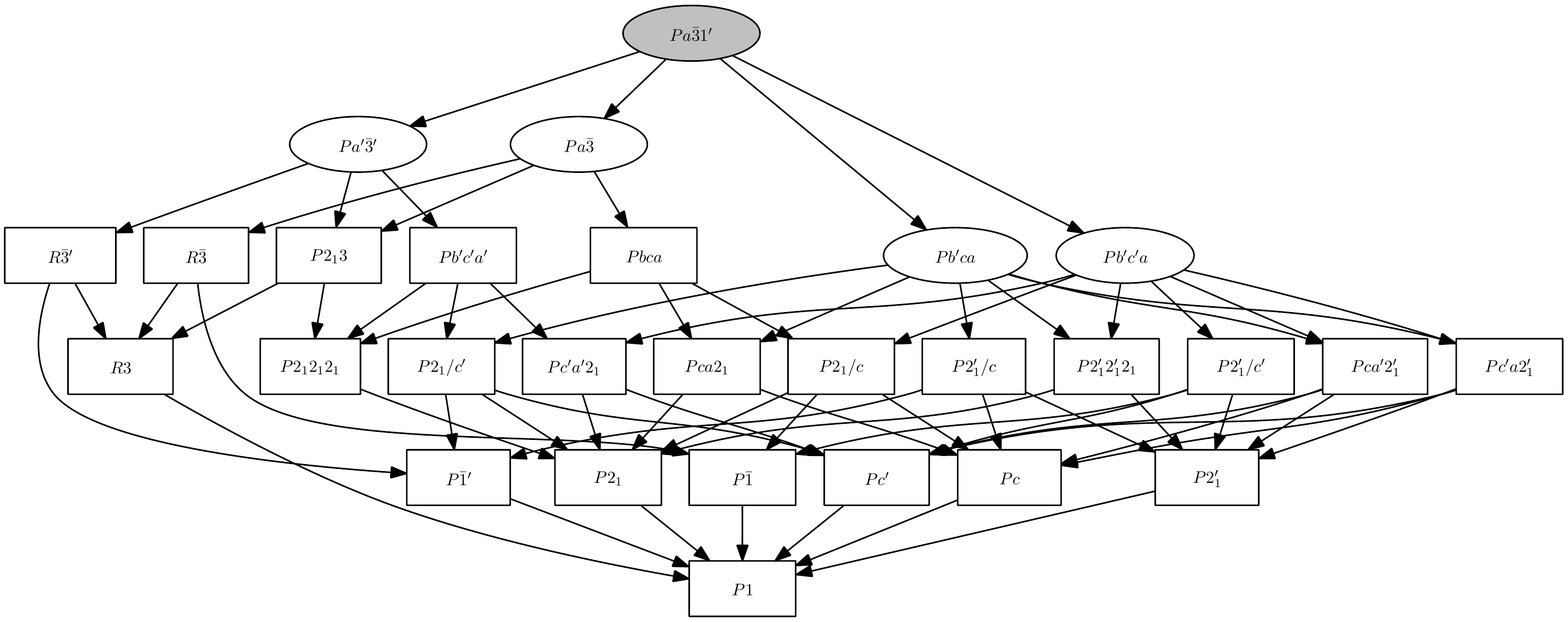}
\caption{Graph of the maximal subgroups of space group $Pa{\bar 3}1'$ (205.34) } 
\label{fig3}
\end{figure*}

As in previous section, we analyze the electronic properties of a magnetic phase in space group $Pa\bar 31'$ by means of magnetic group theory. The magnetic symmetry analysis was performed considering only the k=0 propagation vector because of the observation of a spontaneous large ferromagnetic moment  in the magnetization data and assuming a single irreducible representation. Since Fe was observed on both, the Pt and Sb sites, both Wyckoff positions (4a and 8c) were allowed to have an ordered magnetic moment  in the symmetry analysis (see Fig. \ref{fig3}). The results are summarized in Table \ref{posi2}; there are three different interesting groups that allow for ferromagnetic phases ,  $Pb'c'a$, $R\bar{3}$ and $P2_{1} '/c'$, all deriving from the mGM4+ irreducible representation with different order parameter directions. These groups also feature two-fold magnetic Weyl nodes in $Pb'c'a$ and $P2_{1} '/c'$. Unfortunately the observed ferromagnetic order does not allow for the existence of three-fold crossings. Three-fold crossings can only appear in non-colinear magnetic structures that maintain the cubic symmetry.

\begin{table}[h!]
\begin{center}
\caption{Possible dimensions of the irreps at R point of the ferromagnetic phases of Fe-doped PtSb$_2$ in the maximal subgroups of $Pa\bar 31'$. High-symmetry labels of the corresponding magnetic group where the band-crossings appear , dimensionality of the irreps and direction of the easy axis in the cubic basis are shown in columns 2, 3 and 4 respectively. }
\label{posi2}
\begin{tabular}{cccl}
\hline
\hline
Magnetic Group & K point & Dim & FM direction\\
 \hline
 $Pb'c'a$ (No. 61.436)  & R & 2 & $\langle100\rangle$ direction\\
 $R\bar{3}$ (No. 148.17) & T & 1 & $\langle111\rangle$ direction\\
 $P2_{1} '/c'$ (No. 14.79) & C & 2& $\{110\}$ plane\\ 
\hline
\hline
\end{tabular}
\end{center}
\end{table}

%

\section{Conclusions}

In conclusion we showed theoretically, using magnetic group theory and ab-initio calculations, that the compound PtFeSb in $P2_{1}31'$ has the potential to host new fermionic excitations in the presence of magnetic order.  
Depending on the exact spin order, two different topological phases can exist, exhibiting either two-fold and three-fold magnetic Weyl nodes. This is the first study providing theoretical evidence for magnetic order being a tool to create three-fold degenerate magnetic new fermions. The implications range far beyond the material candidate introduced here, further magnetic compounds in SG $P2_{1}3$ will potentially show similar features, since our arguments are solely based on group theory and not depend on the individual elements composing the material. This formalism could for example also be applied to Mn$_3$IrSi\cite{PhysRevB.69.054422} reported to crystallize in SG $P2_{1}3$.
Our efforts of synthesizing PtFeSb resulted in a different crystallographic phase, a magnetic alloy in SG $Pa\bar 31'$  (No. 205.34). We showed that this phase can also hold two-fold magnetic Weyl fermions, making this material interesting for future investigations as well. At this point, we can't exclude that PtFeSb can also be synthesized in its ordered version. More synthesis methods, maybe at elevated pressures, to avoid formation of the Fe-Pt alloy, might be necessary to synthesize this phase. Nonetheless, both phases, the disordered and ordered version are of high interest for future investigation with spectroscopic methods to confirm the presence of magnetic Weyl nodes.

\section{Acknowledgements} 
MGV was supported by IS2016-75862-P national project of the Spanish MINECO. The work of LE was supported by the Government of the Basque Country (project IT779-13) and the Spanish Ministry of Economy and Competitiveness and FEDER funds (project MAT2015-66441-P).
This research was partially supported by NSF through the Princeton Center for Complex Materials, a Materials Research Science and Engineering Center DMR-1420541.
\bibliography{main3}

\end{document}